\documentclass[pdflatex,sn-mathphys-num]{sn-jnl}% Math and Physical Sciences Numbered Reference Style
\usepackage{graphicx}%
\usepackage{multirow}%
\usepackage{amsmath,amssymb,amsfonts}%
\usepackage{amsthm}%
\usepackage{mathrsfs}%
\usepackage[title]{appendix}%
\usepackage{xcolor}%
\usepackage{textcomp}%
\usepackage{manyfoot}%
\usepackage{booktabs}%
\usepackage{algorithm}%
\usepackage{algorithmicx}%
\usepackage{algpseudocode}%
\usepackage{listings}%
\usepackage{url}
\usepackage{lineno}
\theoremstyle{thmstyleone}%
\theoremstyle{thmstyletwo}%

\theoremstyle{thmstylethree}%

\begin{document}

% \title[Article Title]{High-resolution Atomic Spectroscopy using a Speckle Spectrometer}
\title[Article Title]{Ultra-high precision speckle spectrometer enabling radio-frequency scale resolution of atomic spectra}
%Speckle spectroscopy with radio-frequency scale resolution
%Speckle spectroscopy with mega-hertz reolsutions
%Ultra-high precision speckle spectrometer enabling RF measurement of atomic spectroscopy
%Ultra-high precision speckle spectrometer enabling radio-frequency scale resolution of atomic spectra

\author[1,2,3]{\fnm{Gabriel} \sur{Britto Monteiro}}
\equalcont{These authors contributed equally to this work.}

\author*[1,2,3]{\fnm{Christopher} \sur{Perrella}}\email{chris.perrella@adelaide.edu.au}
\equalcont{These authors contributed equally to this work.}

\author[2,3]{\fnm{Sarah K.} \sur{Scholten}}

\author[4]{\fnm{Morgan} \sur{Facchin}}

\author[2,3]{\fnm{Andre N.} \sur{Luiten}}

\author[4]{\fnm{Graham D.} \sur{Bruce}}

\author*[1,2,3,4]{\fnm{Kishan} \sur{Dholakia}}\email{kishan.dholakia@adelaide.edu.au}

\affil*[1]{\orgdiv{Centre of Light for Life, School of Biological Sciences}, \orgname{Adelaide University}, \orgaddress{\street{North Terrace}, \city{Adelaide}, \postcode{5005}, \state{SA}, \country{Australia}}}

\affil[2]{\orgdiv{Institute for Photonics and Advanced Sensing (IPAS)}, \orgname{Adelaide University}, \orgaddress{\street{North Terrace}, \city{Adelaide}, \postcode{5005}, \state{SA}, \country{Australia}}}

\affil[3]{\orgname{ARC Centre of Excellence in Optical Microcombs for Breakthrough Science (COMBS)}}

\affil[4]{\orgdiv{SUPA, School of Physics and Astronomy}, \orgname{University of St Andrews}, \orgaddress{\street{North Haugh}, \city{St. Andrews}, \postcode{KY16 9SS}, \state{Scotland}, \country{UK}}}


\abstract {Laser speckle, the granular intensity pattern arising from random optical interference, provides a high-dimensional encoding of spectral information that can be exploited for precision metrology. Speckle-based spectrometers have advanced rapidly owing to their compact footprint, mechanical robustness and alignment agnostic nature, yet their spectral resolution has remained limited to the picometre scale. In this work, we break this limit by employing an integrating sphere as a multiply scattering cavity with access to a high range of path lengths to enhance spectral sensitivity. At 780$\,$nm, the resulting device achieves a resolution of 6$\,$fm, corresponding to a resolving power of $1.3\times10^8$, representing an approximately 80-fold improvement over previous implementations. This ultra-high resolution enables clear discrimination of laser sidebands generated by an electro-optical modulator, with extracted sideband powers agreeing with expected values to within 1\%. It further permits the first direct speckle-based measurement of the hyperfine structure of the D\textsubscript{2} transition in \textsuperscript{85}Rb, with transmission spectra differing by no more than 3.6\% from independent wavemeter-referenced measurements. These results establish speckle as a new platform for ultra-high precision spectroscopy, radio-frequency spectrometry, and microwave photonics.}




\maketitle

\section{Introduction}\label{sec1}

Laser speckle is a complex two-dimensional intensity pattern that arises from the interference of electromagnetic waves with randomly distributed phases.
Commonly, speckle patterns are associated with a loss of information and are perceived as contributors to noise in optical systems.
However, the apparent random nature of the speckle is deterministic and rich in information about: the laser, the scattering medium producing the speckle, and the environment within which the scattering medium is situated.
Understanding the deterministic processes that produce the speckle patterns allows this rich source of information to be utilized.
Unsurprisingly, this understanding has been exploited within imaging fields to mitigate speckle noise~\cite{Bianco2018, Cao2019, Cao2022}, for imaging around corners~\cite{Katz2014, Faccio2020} and deep inside tissue~\cite{Yoon2020, Bertolotti2022}, as well as to correct for air turbulence effects in astronomical images~\cite{Labeyrie1970, Bates1982}.
Beyond imaging applications, speckle has found use in: optical computation and information processing~\cite{Gigan2022, Fei2024}, monitoring patient movement~\cite{WarrenSmith2022}, and assessing blood flow at depth~\cite{Briers2013}.

Perhaps counter-intuitively, speckle patterns have emerged as a high-precision metrology tool with a resolution dependent upon the scatterer that is used to generate them.
For example, displacement between macroscopic objects has been measured to picometre precision~\cite{Facchin2023}. Separately, the refractive index of a gas has been measured to parts-per-billion~\cite{Facchin2022}.
In addition, speckle has been used to make high-precision measurements of laser parameters.
Examples include measurement of the polarisation state of a laser's electric-field to within 5\% of its expected value~\cite{Facchin2020}, and measurement of the azimuthal index of a Laguerre-Gaussian beam, possessing orbital angular momentum, to $2\times10^{-5}$, exceeding the previous best measurements by three orders of magnitude~\cite{perrella_twisted_2025}.
Additionally, the wavelengths of monochromatic lasers have been determined up to attometre precision~\cite{Bruce2019, Bruce2020, Kelley2024}.

%An important area where speckle holds promise for impact is in spectrometry.
Traditional interferometers, which are regularly used for wavelength measurement, face challenges in distinguishing the behaviour of multiple laser lines due to their periodic output. Naively, it seems this would also apply to speckle-based measurements. However, the multi-dimensional nature of speckle makes it possible to resolve the contribution from multiple spectral components to a given speckle pattern. 
%Spectroscopy with speckle is a natural advances of the ability of speckle patterns to be used as a high-precision metrology tool and the high precision wavelength measurements demonstrated.
In fact, both reconstructive and speckle spectrometers have proliferated across free-space~\cite{Malone2023, Wang2014, deng_compact_2024, french_speckle-based_2017, kohlgraf-owens_transmission_2010, wang_high_2023, Yang2017}, multi-mode fibre~\cite{Coluccelli2016, Redding2014,Redding_allfiber_2013, chen_high_2023, liang_polarization_2024, liang_high-resolution_2025, liew_broadband_2016, meng_multimode_2019, tan_optical_2023, cen_microtaper_2023, Kurekci2023, wan_high-resolution_2015, redding_using_2012}, and photonic-chip~\cite{Yao2023, Redding2016, Zhang2022, Redding2013, guo_high-resolution_2024, kumar_performance_2025, lee_reconstructive_2025, Lin2023, sun_compact_2022, tian_single-shot_2025, zhang_scalable_2025, zhang_camera-free_2025, hartmann_broadband_2020, piels_compact_2017, li_inverse-designed_2023, Yi2021, yi_high-resolution_2022, zhang_compact_2021, zhang_miniature_2024} platforms.
Despite the effort, the spectral resolutions of these works have not progressed beyond picometre resolution~\cite{Coluccelli2016,Uttam2020}. Increasing the resolution of speckle spectrometers towards femtometre precision would open up potential use in MHz resolved radio-frequency spectrometry as well as metrology applications such as atomic and molecular spectroscopy, and potentially frequency standards for the first time.

%intuitivly it seems difficult to disentangle multiple wavelengths, however this is possible with such granular patterns.
%Previous applications dealt with a single laser line, we 
In this paper, we present a step-change in resolving power to the speckle spectrometer through the use of an integrating sphere as the scattering medium.
An integrating sphere produces multiple scattering events within its volume, producing a wide range of path lengths that interfere at its output, which is crucial when measuring parameters for which the speckle pattern depends upon accrued phase~\cite{Facchin2024}.
%This is critical for a wavelength measurement and allows integrating spheres to outperform photonic chip and multimode fibre spectrometers~\cite{Facchin2024}.
Here we show that a speckle spectrometer based on an integrating sphere achieves 6\,fm wavelength resolution at 780\,nm, which represents an approximate 80-fold improvement over previous speckle based measurements in terms of resolving power~\cite{Coluccelli2016}.
Further, we demonstrate this speckle spectrometer's ability to simultaneously measure up to eleven optical modes, with a power accuracy of 0.2\,\% relative to the total optical power.
To demonstrate the exceptional properties of this spectrometer, we measure the saturated absorption spectra of rubidium.
Rubidium is the primary element underpinning much of modern quantum technology, including chip-scale atomic clocks~\cite{Knappe2004}, neutral-atom quantum processors~\cite{Ebadi2021}, and high-precision quantum gravimeters~\cite{Peters1999}.
We show that the spectrometer can perform spectroscopy of the $5S_{1/2}\,{\rightarrow}\,5P_{3/2}$ transition of $^{85}\text{Rb}$ and resolve the hyperfine structure of the transition.
This demonstrates high precision atomic spectroscopy using speckle for the first time and is a step to wider adoption of speckle spectroscopy across a number of domains.
%The high wavelength resolution and power accuracy of this spectrometer elevates speckle spectrometers to potential use in radio-frequency spectrometry as well as metrology applications such as atomic and molecular spectroscopy, high-resolution frequency comb spectroscopy, and potentially frequency standards for the very first time.

\section{Results}\label{sec2}
\subsection{Experimental setup and calibration}\label{subsec:setup_and_calib}
The heart of the speckle spectrometer demonstrated here is an integrating sphere, which is a hollow spherical cavity with a highly reflective matte surface and two apertures that are used for coupling light in and out of the sphere. An integrating sphere is used as it produces a high-variance distribution of optical path lengths between its input and output. This large variance results in a speckle pattern whose spatial intensity profile is very sensitive to changes in wavelength of the input light~\cite{Facchin2021, Facchin2024}. 

\begin{figure}[t]
    \centering
    \includegraphics[width=\textwidth]{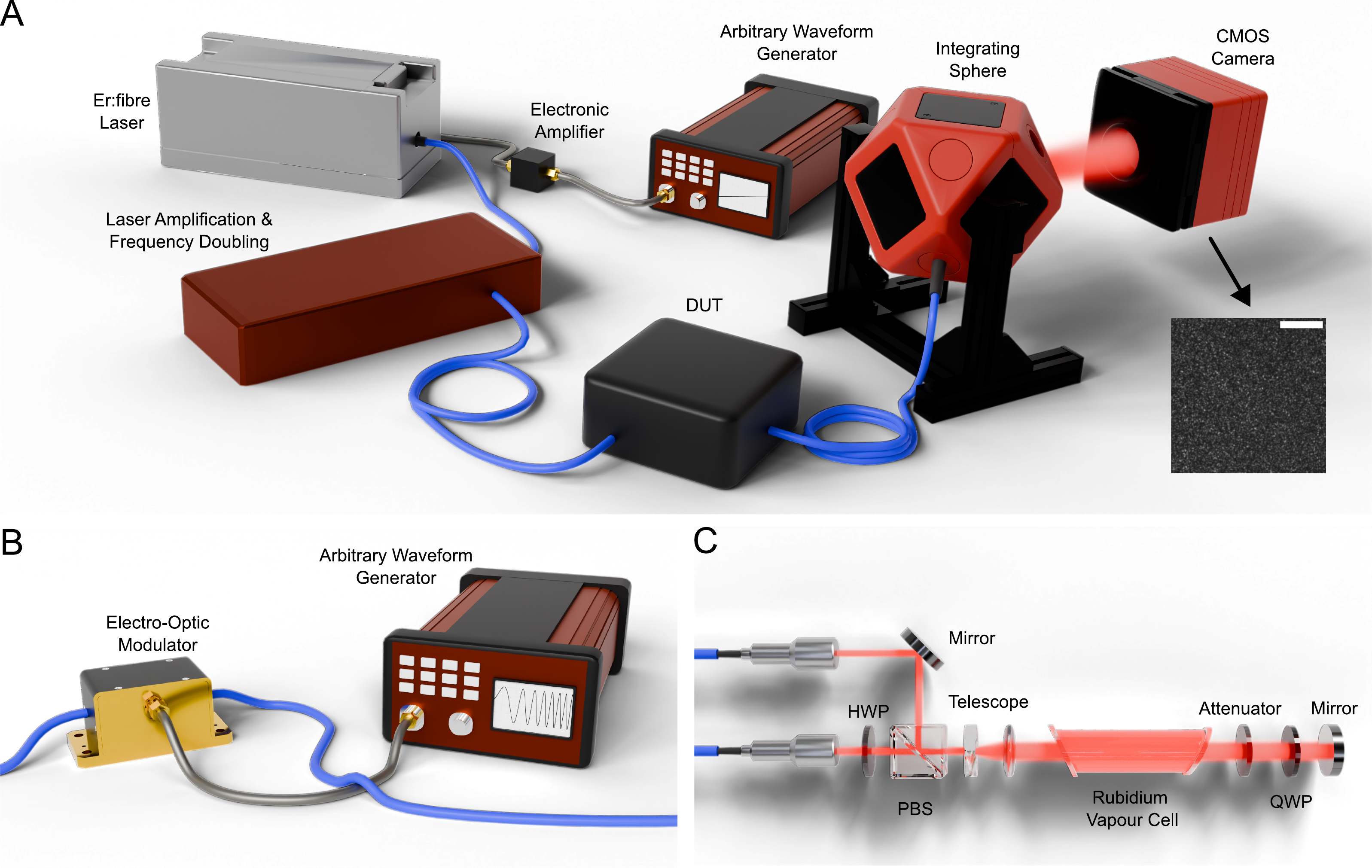}
    \caption{\textbf{A)} Experimental setup: an amplified Er:fibre laser is frequency doubled before passing through the device under test (DUT) and being sent to the integrating sphere. Laser speckle is generated by the integrating sphere and is then captured on a CMOS camera. A cropped speckle pattern of $200\times200$ pixels is shown below the camera as an example. The scale bar is 500\,\textmu{m}. Two DUTs were employed. \textbf{B)} The first DUT: a fibre-coupled electro-optic modulator which was driven by an arbitrary waveform generator. \textbf{C)} The second DUT: rubidium saturation absorption spectroscopy setup (see Methods \autoref{subsec:experimental_app}). PBS: polarisation beam splitter, HWP: half-wave plate, QWP: quarter-wave plate.}
    \label{fig:ExpDiag}
\end{figure}

To test the integrating sphere based speckle spectrometer, the experimental setup shown in Figure~\ref{fig:ExpDiag} was used.
An amplified Er:fibre laser generated approximately 0.7\,W of light around 1560.5\,nm, with a linewidth of $<1\,\text{kHz}$ measured over 120\,\textmu\text{s}.
In free-space, the laser was passed through a second harmonic generation crystal to generate 780\,nm light.
The frequency doubled light was then coupled into polarisation maintaining (PM) fibre and directed towards the device under test (DUT).

The measurement process of the spectrometer consisted of two steps: firstly, a calibration step, and secondly, a measurement step. The calibration of the spectrometer assumed that a given single-wavelength input spectrum produced a unique speckle pattern. This allowed a speckle image, $I$, to be expressed as a linear transformation of an input spectrum, $S$, through a matrix $T$ (i.e., $I = T\times S)$~\cite{Hang2010, Redding_allfiber_2013}. Constructing $T$ experimentally corresponded to calibrating the spectrometer. To this end, speckle patterns at different wavelengths of interest were captured by a CMOS camera as the laser frequency was swept linearly. Each of the patterns captured during the wavelength modulation was flattened into a vector of pixel values and formed a column of $T$. During this process, the wavelength of the laser was recorded by an external wavemeter for reference. Following calibration, the DUT may be interrogated. An unknown input spectrum may be reconstructed using a measured speckle pattern and the inverse of $T$ via: $S = T^{-1}\times I$. In general, $T$ is not square and a pseudo-inverse was used instead (see Methods \autoref{subsec:spectrom_calib})~\cite{Hang2010, Redding_allfiber_2013, Redding_noise_2014, Redding2016}
The vector, $S$, contains the proportion of power at each calibration wavelength in the pattern $I$, thereby corresponding to the optical spectrum.

\subsection{Radio-frequency spectrometry}\label{subsec:rf_spec}
The first DUT was an electro-optic modulator (EOM), see \autoref{fig:ExpDiag}B, which was used to test the spectral and optical power resolution limits of the spectrometer. This was achieved by sinusoidally modulating the phase of the laser's electric field ($E(t)$), thereby creating sidebands of varying power and frequency. %Sinusoidal phase modulation of a wave is described mathematically by the Jacobi–Anger expansion~\cite{Korsch2006}:
This is described mathematically by the Jacobi–Anger expansion~\cite{Korsch2006}:
\begin{equation}\label{eqn:JacobiAnger}
    E(t) = A \cos\left[ 2 \pi f_c t + \beta \sin\left( 2 \pi f_m t\right) \right] = A \sum_{n=-\infty}^\infty J_n(\beta) \cos\left[ 2 \pi (f_c+n f_m) t\right]
\end{equation}
where $A$ is amplitude of the modulated waveform (i.e., the amplitude of the electric field), $f_c$ is the frequency of the carrier (i.e., the optical frequency of the laser), $\beta$ is the applied phase modulation depth, $f_m$ is the frequency of the sinusoidal phase modulation, and $J_n(\beta)$ are Bessel functions of the first kind.
Light was coupled into and out of the EOM and into the sphere using PM fibre, thereby ensuring a constant output spatial mode. The output of the fibre was uncollimated such that a large area of the sphere's inner surface was illuminated. After numerous internal scattering events, light exited the sphere via its second aperture, 5\,mm in diameter, and produced the output speckle patterns which were used for measurement. The output patterns were imaged 60\,mm from the output aperture, thereby yielding grains that were approximately two pixels wide in the horizontal and vertical directions, and ensuring speckle grains were not spatially under sampled. A CMOS camera with a resolution of $1600\times1100$ pixels was used to image the speckle patterns.

\begin{figure}[t]
    \centering
    \includegraphics[width=\linewidth]{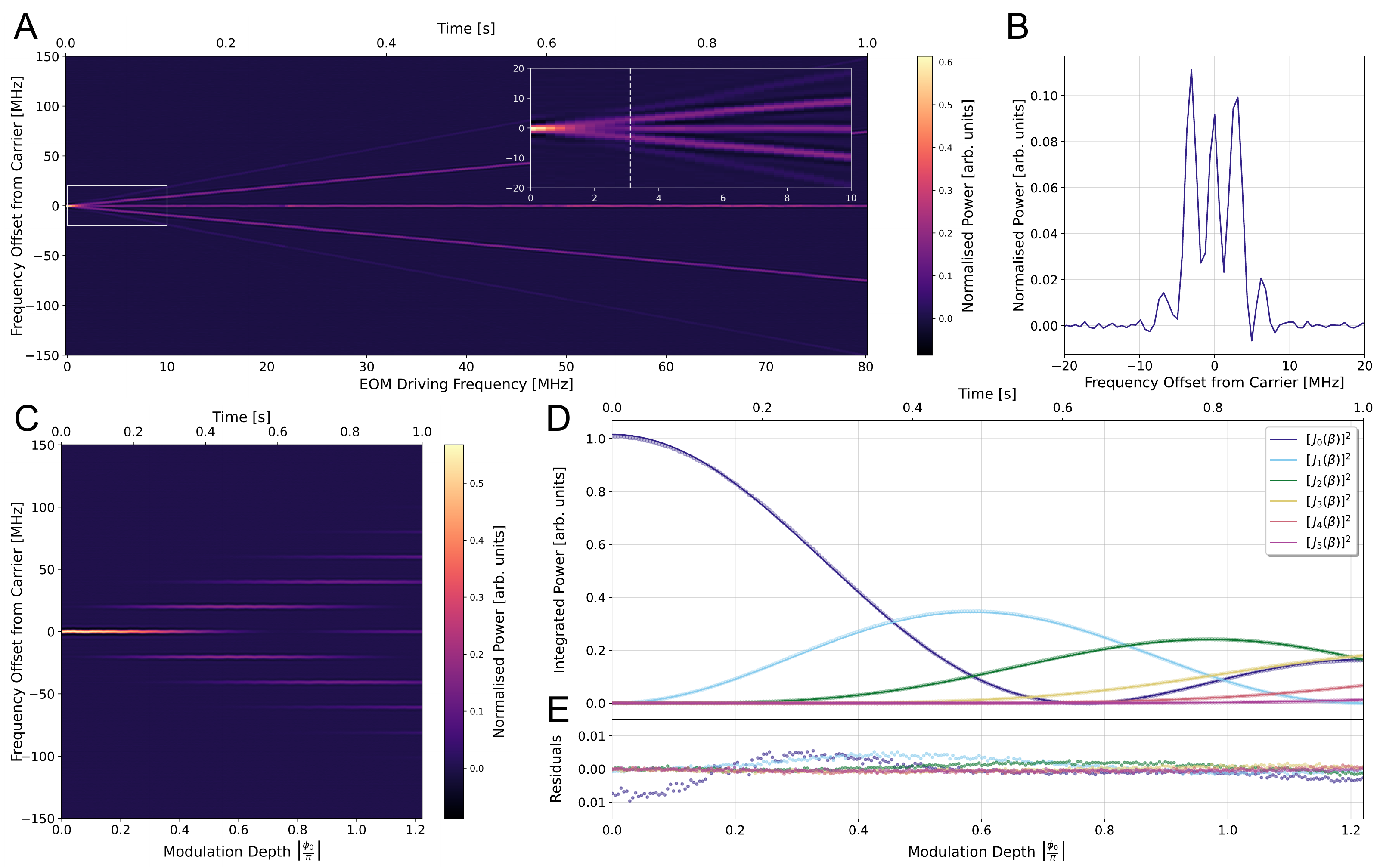}
    \caption{Integrating sphere speckle spectrometer characterisation using an electro-optic modulator (EOM) as the device under test. \textbf{A)} Spectrogram obtained as the frequency of the sinusoid driving the EOM was swept linearly from approximately 0\,Hz to 80\,MHz in one second. In this dataset, the amplitude of the driving sinusoid was held constant. The inset is a zoom into the region where the EOM driving frequency lies between $\sim0\,\text{Hz}$ and 10\,MHz. The vertical dashed line at approximately 3\,MHz shows the EOM driving frequency for which the laser carrier and sidebands are resolved. \textbf{B)} Individual reconstructed spectrum from \textbf{A} corresponding to when the EOM driving frequency was 3\,MHz. \textbf{C)} Spectrogram obtained as the amplitude of the sinusoid driving the EOM was modulated linearly from 0\% to 100\%. At a modulation depth of approximately 1.19, the spectrometer is able to resolve eleven separate lines ($\pm5$ sidebands and one carrier). \textbf{D)} Numerically integrated power in the laser carrier and sidebands as a function of phase modulation depth. The power in the $\pm n^{\text{th}}$ sidebands was averaged in each spectrum and is presented as a single data point. Solid lines are fits to the experimental data using squared Bessel functions of the first kind. The fit is performed for all functions simultaneously to ensure parameters do not change between functions. \textbf{E)} Residuals between numerically integrated powers and squared Bessel functions in \textbf{D} as a function of phase modulation depth.}
    \label{fig:eomCharacterisation}
\end{figure}

%\autoref{fig:eomCharacterisation}A shows spectra reconstructed from the integrating sphere speckle. Here the EOM was driven by a sinusoid with a constant amplitude of 3.75\,V peak-to-peak as this produced a spectrum with approximately equal power in the carrier and $\pm1^{\text{st}}$ order sidebands.

\autoref{fig:eomCharacterisation}A shows spectra reconstructed from the integrating sphere speckle. Here the EOM was driven by a sinusoid with a constant amplitude which was chosen to produce an output spectrum with approximately equal power in the carrier and $\pm1^{\text{st}}$ order sidebands. At this modulation depth, approximately 4.5\% of the optical power is in the $\pm2^{\text{nd}}$ order sidebands, while higher orders contain negligible power. The frequency of the driving sinusoid was swept linearly from approximately 0\,Hz to 80\,MHz in one second. This was done to produce spectra with laser sidebands which had increasing frequency offsets from the laser carrier, as seen in \autoref{fig:eomCharacterisation}A. Since the frequency of the waveform that was driving the EOM was known at each time step, the spectrogram was used to estimate the spectral resolution of the spectrometer. This is defined here as the minimal frequency separation between infinitely narrow peaks of the same power, which yielded separate peaks after reconstruction (i.e., similarly to the Rayleigh Criterion~\cite{Rayleigh1879}). This definition of resolution is chosen to highlight the difference between the function of a spectrometer and that of a wavemeter. In the wavemeter's case, the test of the instrument is its ability to reliably measure the position of a laser's spectrum along a spectral axis. For the spectrometer, the test shifts to the device's ability to separate different closely spaced spectral features. In these terms, our spectrometer possesses a resolution $<3\,\text{MHz}$, or 6\,fm at a wavelength of 780\,nm. This is in good agreement with the half-width-at-half-maximum of the Pearson correlation curve (4.4MHz) which may also be used to estimate the resolution~\cite{Facchin2024}. The spectrum obtained for the EOM driving frequency where the laser sidebands are resolved is shown in \autoref{fig:eomCharacterisation}B.

After the spectral resolution of the spectrometer was determined, the power accuracy was investigated. To this end, light passing through the EOM was instead modulated using a sinusoid with a constant frequency ($f_m=\text{20\,MHz}$), and a linear amplitude modulation, from 0\% to 100\%, over a period of 1 second. The increase in the amplitude of the driving waveform corresponded to an increase in modulation depth. \autoref{eqn:JacobiAnger} describes the resulting amplitude modulation of the laser carrier and sidebands, which are modulated by Bessel functions of the first kind. Speckle patterns captured during this period were reconstructed into spectra and compiled into the spectrogram shown in \autoref{fig:eomCharacterisation}C. \autoref{fig:eomCharacterisation}D shows the power of the laser carrier and sidebands as a function of phase modulation depth. To account for spread across multiple spectral channels, the power was calculated by numerically integrating the area under each of the peaks. The power in each spectral feature was fitted to its respective squared Bessel function. Squared functions are used since \autoref{eqn:JacobiAnger} describes the behavior of the electric field and the laser speckle images measure optical intensity. Residuals are plotted in \autoref{fig:eomCharacterisation}E. The fits show excellent agreement with the measured sideband powers: at worst the data deviated from the expected power by approximately 1\,\%, overall the standard deviation of the residuals was approximately 0.2\,\% across all sidebands.

\begin{figure}[t]
    \centering
    \includegraphics[width=\linewidth]{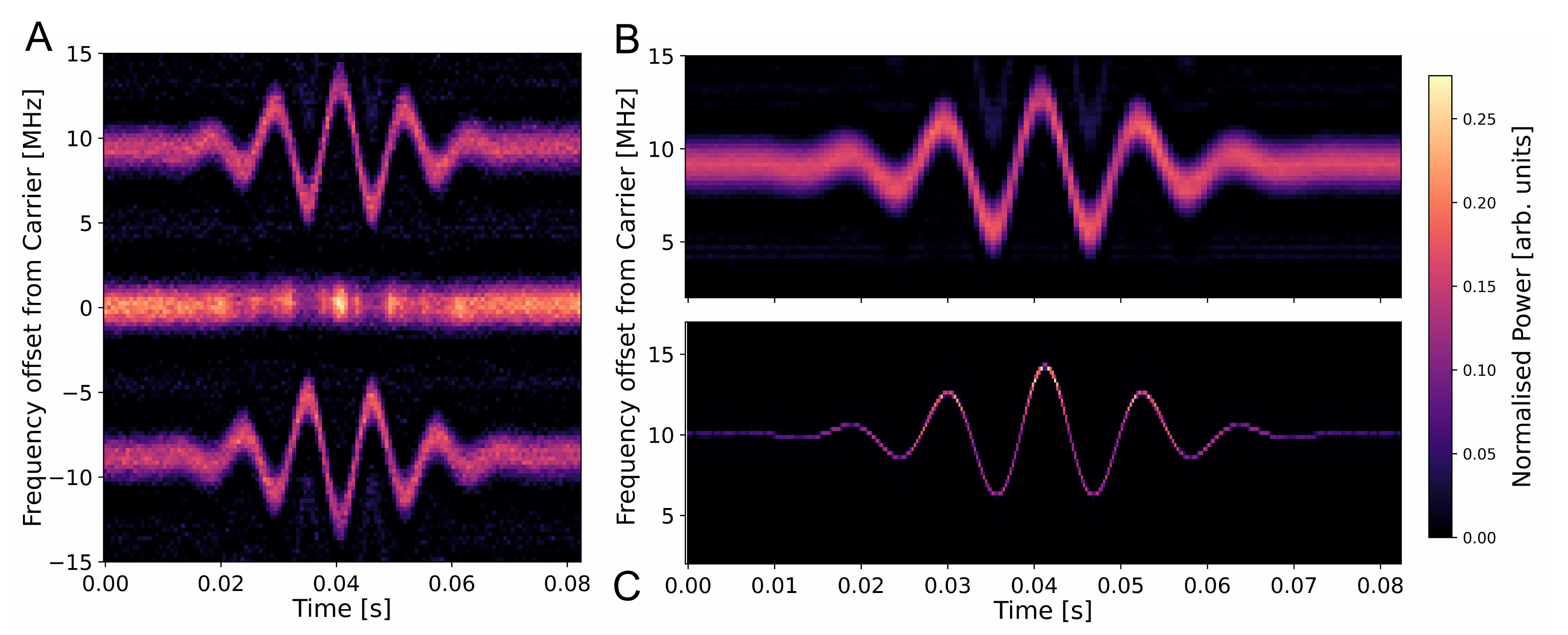}
    \caption{\textbf{A)} Spectrogram obtained as the EOM was driven by a sinusoid which had its frequency and amplitude modulated to produce spectra resembling wave packets. This figure shows one period of the modulation with no averaging. \textbf{B)} Zoom into the $+1^\text{st}$ order sideband of the spectrogram obtained after averaging over twenty modulation periods. \textbf{C)} Spectrogram of the expected behaviour of the $+1^\text{st}$ order sideband given the applied modulation to the sinusoid which drove the EOM.}
    \label{fig:wavepacket}
\end{figure}

To demonstrate the spectrometer's ability to reconstruct complex waveforms, the EOM was driven with a waveform which had its frequency and amplitude modulated to produce sidebands whose temporal behaviour resembled wave packets. \autoref{fig:wavepacket}A shows the spectrogram obtained after one period of the modulation on the driving waveform, \autoref{fig:wavepacket}B shows an averaged spectrogram after twenty modulation periods, and \autoref{fig:wavepacket}C shows the expected spectrogram given the modulation which was applied. The periodicity and phase of the sideband modulation is well-preserved in the reconstructed spectra, while the visibly increased width of the peaks in the experimental spectra is reflective of the spectrometer’s resolution of approximately 3 MHz. In \autoref{fig:wavepacket}A \& B it is possible to see a faint trace of the $\pm2^\text{nd}$ order sidebands, though they possess negligible power. 

%Specifically, the EOM was driven with a sinusoid which had a sinusoidal frequency modulation and a gaussian amplitude envelope. 

\subsection{Atomic spectroscopy}

To test the capability of the integrating sphere speckle spectrometer for potential use in atomic and molecular spectroscopy, the DUT is swapped from the EOM to a rubidium (Rb) saturated absoption spectroscopy (SAS) setup. This was used to measure a series of hyperfine absorption features of rubidium-85 (Rb-85). The Er:fibre laser wavelength was tuned to the $5S_{1/2}\,{\rightarrow}\,5P_{3/2}$ transition of Rb-85, known as the D$_2$ transition~\cite{steck2025}.
The calibration of the speckle spectrometer was performed as previously, except that the Rb SAS (\autoref{fig:ExpDiag}C) setup was by-passed during the process (see Methods \autoref{subsec:experimental_app}). During calibration, the Er:fibre laser frequency was swept approximately 1.6\,GHz over 5.33\,s. After this period, light was injected into the SAS setup using PM fibre for measurement of the D$_2$ transition. The laser frequency was then swept again in the same manner over a reduced frequency range (1.28\,GHz). This was done to ensure laser drift would not introduce artefacts near the edges of the test spectra. 

\begin{figure}[t]
    \centering
    \includegraphics[width=\linewidth]{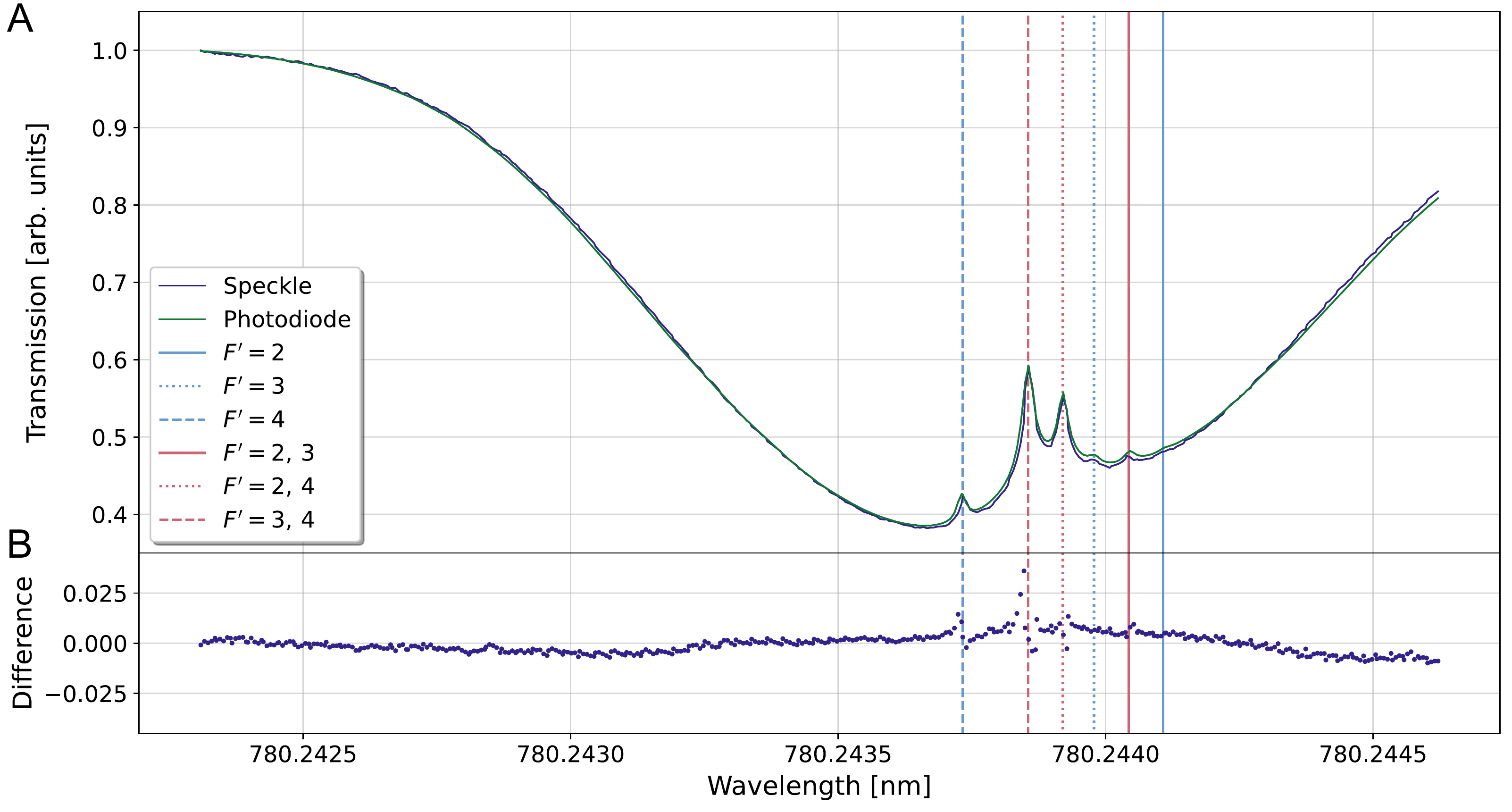}
    \caption{Saturated absorption spectrum of the Rb-85 D$_2$ transition. \textbf{A)} Comparison of the D$_2$ transition as captured by the integrating sphere speckle spectrometer and a photodiode. The photodiode trace was captured at a much higher sampling rate and has been decimated, averaged, and interpolated to match the sampling rate and wavelength of the speckle measurement. Blue vertical lines show the frequencies of the transitions from the $5S_{1/2}$, $F=3$ state to the $5P_{3/2}$, $F^{\prime}=\text{2, 3, and 4}$ states~\cite{steck2025}. Red vertical lines lines show the frequencies of the crossover transitions. The wavelengths for both the speckle and photodiode traces were measured separately using a reference wavemeter and then calibrated using the known $F=3\,\rightarrow\,F^{\prime}=4$ transition. \textbf{B)} Difference between the speckle and photodiode SAS traces in \textbf{A}.}
    \label{fig:RbSpectra}
\end{figure}

Spectral reconstruction yielded a set of spectra with a single peak. The power in these peaks was integrated numerically, as in \autoref{subsec:rf_spec}, and the wavelength of the laser was determined by fitting the peaks in each spectrum with a Gaussian and extracting the centre wavelength from the fit parameters. These two values were then plotted against each other and displayed in \autoref{fig:RbSpectra}A. For comparison, a measurement of the D$_2$ transition that was recorded using a photodiode and wavemeter is also presented. The difference between the speckle spectrometer trace and the photodiode trace were at maximum 3.6\,\%, overall the standard deviation of the difference was 0.5\,\%.

%%%%%%%%%%%%%%%%%%%%%%%%%%  Discussion  %%%%%%%%%%%%%%%%%%%%%%%%%%
\section{Discussion}
Despite the proliferation of speckle spectrometers, the wavelength resolution has only reached the picometre level in both multi-mode fibre spectrometers~\cite{Coluccelli2016}, and photonic chip/chip-scale spectrometers~\cite{Uttam2020}. In contrast, the speckle spectrometer presented in this work, based on an integrating sphere, possesses a resolution of $<3$\,MHz or 6\,fm at 780\,nm. A useful way of quantifying the capability of a spectrometer, is the resolving power, $R$, which can be defined as~\cite{Kurekci2023}:
\begin{equation}\label{eqn:resolvingPower}
    R = \frac{\lambda}{\delta\lambda}
\end{equation}
where $\lambda$ is the operating wavelength, and $\delta\lambda$ is the spectral resolution. Conventional spectrometers (i.e., those based upon dispersive elements) can achieve resolving powers of $10^4$ to $10^5$~\cite{Kurekci2023,wan_high-resolution_2015}. In contrast, the highest resolving power speckle spectrometers in literature report values of approximately $1.5\,{\times}\,10^6$~\cite{Coluccelli2016, Uttam2020, Redding2014}. 
In comparison, our spectrometer possesses a resolving power of approximately $1.3\times10^8$. 
This 80-fold improvement, when compared to previous works, is due to the choice of scattering medium. Inside the integrating sphere light may take many possible paths and will scatter multiple times before exiting, producing a wide distribution of path lengths between the input and output. This leads to a speckle pattern whose spatial intensity profile is highly sensitive to path-dependent perturbations, such as wavelength changes~\cite{Facchin2024, goodmanSpeckle}.

The downside of such sensitivity is its lack of discrimination. Speckle generated by the integrating sphere is extremely sensitive to all path dependent perturbations. Such unwanted fluctuations, commonly referred to as drift, can arise from temperature changes, mechanical deformation of the scatterer, or variations in refractive index~\cite{Facchin2024}. In practice, this means that the calibration of the spectrometer is valid only for a limited duration. We estimated this validity period as approximately 60\,s by recording speckle patterns under nominally unperturbed conditions and determining when the Pearson correlation between an initial pattern and later patterns reduced to 0.5~\cite{Facchin2024}. Although the drift is significant, it did not affect our measurements, which were performed on 5–10\,s timescales. 
We note that most devices with this calibre of wavelength/frequency resolution require significant engineering to maintain their stability in the long term. For example, high precision optical devices typically employ thermal engineering utilising low thermal coefficient materials, passive isolation from the environment, and active temperature control. Examples of such devices include ultra-low expansion optical cavities~\cite{alnis_subhertz_2008, davila-rodriguez_compact_2017}, fibre interferometers~\cite{kefelian_ultralow-frequency-noise_2009, dong_subhertz_2015}, and atomic transitions used in optical clocks~\cite{Ludlow2015}. 
Employing such techniques will extend the validity period of the integrating sphere spectrometer. For example, active temperature control and vacuum isolation from the environment have been shown to increase the calibration stability to more than three hours in multi-mode fibre speckle spectrometers~\cite{Coluccelli2016}. 

Speckle spectrometers with high spectral resolution, such as the 3\,MHz resolution demonstrated here, will find use in many applications:
Laser frequency stabilisation is a key area of interest, for interferometry~\cite{kefelian_ultralow-frequency-noise_2009, dong_subhertz_2015}, tests of fundamental physics~\cite{katori_optical_2011, LIGO_2015, Ludlow2015, safronova_search_2018, Grotti2018}, and precision measurements~\cite{diddams_optical_2020}.
This spectrometer, with the aforementioned improvements to its validity period, would allow stabilisation of a laser to an arbitrary frequency~\cite{metzger_harnessing_2017}.
This would greatly simplify opto-electronic setups that form the basis of magneto-optical traps and Bose-Einstein condensates.
These experiments typically require multiple lasers stabilised to atomic transitions, often spectrally detuned from each other by radio-frequencies~\cite{schreck_laser_2021}. This is especially the case when cold atoms need to be transported between locations using dipole traps~\cite{Hilton2018a}, or in quantum computing applications that individually address atoms in a lattice~\cite{chiu_continuous_2025}. 
This spectrometer could replace multiple radio-frequency (RF) phase locks in these setups.

%The radio-frequency scale resolution of the integrating sphere speckle spectrometer enables direct RF spectrometry in the optical domain, relieving the necessity for homodyne and/or heterodyne detection. Consequently, this spectrometer could find applications in microwave photonic scenarios, in areas such as communications, radar, astronomy, sensing and biomedicine~\cite{Marpaung2019, RomeroCortes2020, Zou2016}.
The radio-frequency scale resolution of the integrating sphere speckle spectrometer enables direct MHz-resolved RF spectrometry in the optical domain. This is relevant to the field of microwave photonics, where the aim is to generate, manipulate, process, transport, and measure RF signals using optical techniques. Microwave photonics promises large instantaneous bandwidths, low loss, and wide frequency coverage, which can advance fields such as: communications, RADAR, astronomy, sensing, and biomedicine~\cite{Marpaung2019, RomeroCortes2020, Zou2016}.
The ability for our spectrometer to measure RF signatures in the optical domain alleviates the requirement for homodyne and/or heterodyne detection to extract these signals, and would reduce the need for multiple steps of RF to optical conversions.

\section{Conclusion}
Employing an integrating sphere as the basis for a speckle spectrometer has yielded unprecedented spectral resolution. Using an electro-optic modulator to generate laser sidebands and speckle from an integrating sphere, we demonstrate a spectral resolution of approximately 3\,MHz. When investigating the power resolution, we demonstrated the spectrometer's capacity to accurately reconstruct the power of laser sidebands to within 1\,\% of expectation whilst simultaneously measuring eleven laser modes. Using the ultra-high spectral resolution of the integrating sphere speckle spectrometer, we measured the hyperfine structure of the D$_2$ transition of rubidium-85. These speckle measurements were comparable to those taken with a photodiode to within 1.5\,\%. These results open the door to applications for speckle in radio frequency spectrometry and atomic spectroscopy.  

\section{Methods}\label{sec:methods}
\subsection{Experimental apparatus}\label{subsec:experimental_app}
The Er:fibre laser used here was an NKT Koheras Adjustik E15. This was then amplified by an erbium doped fibre amplifier (NKT Koheras Boostik HPA) and frequency doubled in free-space using a second-harmonic generation crystal (Covesion periodically poled lithium niobate) before being coupled into PM fibre. This light was split in two (95:5 ratio) using an in-fibre filter-coupler (AFW Technologies PFC-78-2-05). The 5\,\% output was directed to a HighFinesse WS7 wavemeter for reference while the 95\,\% output was sent to the DUTs. The first DUT was an ixBlue (now Exail) EOM (NIR-MPX800-LN-10) which was driven directly by a Keysight 33611A arbitrary waveform generator (AWG). During calibration, the laser frequency was swept over a range of approximately 430\,MHz in five seconds. At this time light still passed through the EOM but it was not driven by the AWG.

The second DUT was a rubidium SAS setup (as seen in \autoref{fig:ExpDiag}C). In the setup, approximately 1.3\,mW of light was launched from PM fibre into free-space using a collimation package (Thorlabs F260APC-780). The polarisation of the light out of the collimator was adjusted using a half-wave plate to maximise the power through a polarisation beam splitter (PBS) cube (Thorlabs CCM1-PBS252/M). The beam was then magnified by 3.5 times using a telescope comprised of two lenses with focal lengths of $-50\,\text{mm}$ and $175\,\text{mm}$. This enlarged beam formed the pump for the SAS setup and was directed through a rubidium cell (Thorlabs CQ19075-RB) before double-passing a $-3\,\text{dB}$ neutral density filter (Thorlabs NE03A) and quarter-wave plate. The attenuated light formed the probe and was polarised orthogonally to the pump beam. The probe beam then passed through the rubidium cell and was reflected by the PBS before being coupled back into PM fibre using another collimation package (Thorlabs F260APC-780). Before launching into the SAS, the light was split further using a 90:10 ratio PM filter-coupler (AFW TechnologiesPFC-78-2-10). The 90\,\% output was used for the SAS while the 10\,\% output was directed towards the integrating sphere for calibration. After calibration, the fibre used for launching into the sphere was swapped from this calibration path to the output of the SAS setup. 

The integrating sphere used here has an inner diameter of 100\,mm, and is made from a PTFE-based bulk material with high reflectance ($>99$\,\%) between 350\,nm and 1500\,nm (Thorlabs 4P3). Light exiting the sphere from its output port was captured on a PhotonFocus MV4-D1600-S01-GT CMOS camera. Before conducting measurements, both the sphere and camera were allowed to equilibrate within a passive acoustic-isolation enclosure while the laser light was injected into the sphere. While characterising the spectral resolution and power accuracy of the spectrometer with the EOM, the camera recorded with an exposure time of 3\,ms and frame-rate of 260\,Hz. When recording the data shown in \autoref{fig:wavepacket}, the camera's region of interest was reduced from $1600\times1100$ to $1600\times60$ pixels, to allow for a faster sampling rate of 1560 frames-per-second with an exposure time of 0.518\,ms. When using the Rb SAS setup as the DUT, an exposure time of 24.9\,ms and frame-rate of 40\,Hz was used. Images were recorded on a computer for analysis. 

\subsection{Spectrometer calibration}\label{subsec:spectrom_calib}
The speckle spectrometer relies on the generation of a matrix $T$ whose columns correspond to output speckle patterns ($I$) at different wavelengths of interest. This matrix was determined by recording the speckle patterns which were produced as the laser wavelength was swept over a region of interest. The wavelength sweep was achieved by tuning the cavity length of the Er:fibre laser using the in-built piezo-electric transducer (PZT). An Agilent 33250A arbitrary waveform generator produced a single period of a triangular wave upon manual triggering, this output was then amplified 7.5 times by a PZT amplifier (Thorlabs MTD694A). The wavelength of the laser during the sweep was recorded by the reference wavemeter (HighFinesse WS7) and stored for analysis. In code, the linear section of the wavelength sweep was fitted. This fit was then used to determine the wavelengths of the speckle patterns used in constructing $T$. As per \autoref{subsec:setup_and_calib}, the inverse of $T$ enables reconstruction of an unknown spectrum from a given speckle pattern. As $T$ was not a square matrix, a pseudo-inverse was used to approximate $T^{-1}$. This pseudo-inverse was determined during analysis by computing a truncated singular value decomposition (SVD). Although multiple publications utilise techniques such as optimised truncations~\cite{Coluccelli2016, Redding_allfiber_2013} or non-linear minimisation~\cite{Redding_allfiber_2013,Redding2013,Redding2016} to improve the accuracy of their spectral reconstruction, no such techniques were employed in any of the work presented here.

\backmatter

\bmhead{Acknowledgements}
The authors acknowledge Alexander Trowbridge for generating the illustrations in \autoref{fig:ExpDiag}. The authors acknowledge Erik P. Schartner for contributions in the establishment of the initial setup. 
The authors acknowledge funding support from an ARC Laureate Fellowship (project number FL210100099).
This research was conducted by the Australian Research Council Centre of Excellence in Optical Microcombs for Breakthrough Science (project number CE230100006) and funded by the Australian Government. This work is supported by the US Office of Naval Research Global (N62909-25-1-2029). Sarah K. Scholten is the recipient of an Australian Research Council Australian Early Career Researcher Industry Fellowship (project number IE240100056) funded by the Australian Government
We thank the Optofab node of the Australian National Fabrication Facility (ANFF) which utilize Commonwealth and South Australia State Government funding. 
%The authors thank Evan Johnson, Alastair Dowler, and Lijesh Thomas and the rest of the Optofab team for their technical support. \textbf{Do we need to thank Optofab?}

\section*{Declarations}

% Some journals require declarations to be submitted in a standardised format. Please check the Instructions for Authors of the journal to which you are submitting to see if you need to complete this section. If yes, your manuscript must contain the following sections under the heading `Declarations':

%\begin{itemize}
% \item Conflict of interest/Competing interests (check journal-specific guidelines for which heading to use)
% \item Ethics approval and consent to participate
% \item Consent for publication
% \item Materials availability
%\end{itemize}

\bmhead{Funding} The authors acknowledge funding support from an ARC Laureate Fellowship (project number FL210100099). This research was conducted by the Australian Research Council Centre of Excellence in Optical Microcombs for Breakthrough Science (project number CE230100006) and funded by the Australian Government. This work is supported by the US Office of Naval Research Global (N62909-25-1-2029). Sarah K. Scholten is the recipient of an Australian Research Council Australian Early Career Researcher Industry Fellowship (project number IE240100056) funded by the Australian Government.

\bmhead{Conflict of interest} The authors declare no conflicts of interest.

\bmhead{Ethics approval and consent to participate} Not applicable.

\bmhead{Consent for publication} The authors consent to publication.

\bmhead{Data availability} The data presented in the findings of this study are available from the corresponding author(s) upon reasonable request.

\bmhead{Materials availability} Not applicable.

\bmhead{Code availability} Not applicable.

\bmhead{Author contribution} K.D., G.D.B, and C.P. conceived the study. G.B.M and C.P. designed and conducted the experiments. M.F., S.K.S., and A.N.L supported the experimental work. G.B.M analysed the data. C.P., G.B.M, and K.D. wrote the manuscript with contributions from other authors. All authors reviewed, edited, and approved the manuscript. K.D. and C.P. supervised the study.

\noindent
%If any of the sections are not relevant to your manuscript, please include the heading and write `Not applicable' for that section. 


\bibliography{sn-bibliography}% common bib file

@article{Bertolotti2022,
	title = {Imaging in complex media},
	volume = {18},
	urldate = {2024-09-14},
	journal = {Nature Physics},
	author = {Bertolotti, Jacopo and Katz, Ori},
	year = {2022},
	pages = {1008--1017},
}

@article{Yoon2020,
	title = {Deep optical imaging within complex scattering media},
	volume = {2},
	journal = {Nature Reviews Physics},
	author = {Yoon, Seokchan and Kim, Moonseok and Jang, Mooseok and Choi, Youngwoon and Choi, Wonjun and Kang, Sungsam and Choi, Wonshik},
	year = {2020},
	pages = {141--158},
}

@article{Katz2014,
	title = {Non-invasive single-shot imaging through scattering layers and around corners via speckle correlations},
	volume = {8},
	journal = {Nature Photonics},
	author = {Katz, Ori and Heidmann, Pierre and Fink, Mathias and Gigan, Sylvain},
	year = {2014},
	pages = {784--790},
}

@article{Briers2013,
	title = {Laser speckle contrast imaging: theoretical and practical limitations},
	volume = {18},
	journal = {Journal of Biomedical Optics},
	author = {Briers, David and Duncan, Donald D. and Hirst, Evan R. and Kirkpatrick, Sean J. and Larsson, Marcus and Steenbergen, Wiendelt and Stromberg, Tomas and Thompson, Oliver B.},
	year = {2013},
	pages = {066018},
}

@article{Faccio2020,
	title = {Non-line-of-sight imaging},
	volume = {2},
	journal = {Nature Reviews Physics},
	author = {Faccio, Daniele and Velten, Andreas and Wetzstein, Gordon},
	year = {2020},
	pages = {318--327},
}

@article{WarrenSmith2022,
	title = {Multimode optical fiber specklegram smart bed sensor array},
	volume = {27},
	journal = {Journal of Biomedical Optics},
	author = {Warren-Smith, Stephen C. and Kilpatrick, Adam D. and Wisal, Kabish and Nguyen, Linh V.},
	year = {2022},
	pages = {067002},
}

@article{Facchin2021,
	title = {Wavelength sensitivity of the speckle patterns produced by an integrating sphere},
	volume = {3},
	journal = {Journal of Physics: Photonics},
	author = {Facchin, Morgan and Dholakia, Kishan and Bruce, Graham D},
	year = {2021},
	pages = {035005},
}

@article{Labeyrie1970,
    title = {Attainment of diffraction limited resolution in large telescopes by Fourier analysing speckle patterns in star images},
    author = {Labeyrie, Antoine},
    journal = {Astronomy and Astrophysics},
    volume = {6},
    pages = {85--87},
    year = {1970}
}

@article{Cao2022,
	title = {Harnessing disorder for photonic device applications},
	volume = {9},
	journal = {Applied Physics Reviews},
	author = {Cao, Hui and Eliezer, Yaniv},
	year = {2022},
	pages = {011309},
}

@article{Gigan2022,
	title = {Imaging and computing with disorder},
	volume = {18},
	journal = {Nature Physics},
	author = {Gigan, Sylvain},
	year = {2022},
	pages = {980--985},
}

@article{Cao2019,
	title = {Complex lasers with controllable coherence},
	volume = {1},
	journal = {Nature Reviews Physics},
	author = {Cao, Hui and Chriki, Ronen and Bittner, Stefan and Friesem, Asher A. and Davidson, Nir},
	year = {2019},
	pages = {156--168},
}

@article{Facchin2023,
	title = {Measuring picometre-level displacements using speckle patterns produced by an integrating sphere},
	volume = {13},
	journal = {Scientific Reports},
	author = {Facchin, Morgan and Bruce, Graham D. and Dholakia, Kishan},
	year = {2023},
	pages = {14607},
}

@article{Facchin2022,
	title = {Measurement of Variations in Gas Refractive Index with $10^{–9}$ Resolution Using Laser Speckle},
	journal = {ACS Photonics},
	author = {Facchin, Morgan and Bruce, Graham D. and Dholakia, Kishan},
	year = {2022},
    volume = {9},
    pages = {830--836}
}

@article{Bruce2019,
	title = {Overcoming the speckle correlation limit to achieve a fiber wavemeter with attometer resolution},
	volume = {44},
	journal = {Optics Letters},
	author = {Bruce, Graham D. and O’Donnell, Laura and Chen, Mingzhou and Dholakia, Kishan},
	year = {2019},
	pages = {1367--1370},
}

@article{Bruce2020,
	title = {Femtometer-resolved simultaneous measurement of multiple laser wavelengths in a speckle wavemeter},
	volume = {45},
	journal = {Optics Letters},
	author = {Bruce, Graham D. and O’Donnell, Laura and Chen, Mingzhou and Facchin, Morgan and Dholakia, Kishan},
	year = {2020},
	pages = {1926--1929},
}

@article{Facchin2020,
	title = {Speckle-based determination of the polarisation state of single and multiple laser beams},
	volume = {3},
	journal = {OSA Continuum},
	author = {Facchin, Morgan and Bruce, Graham D. and Dholakia, Kishan},
	year = {2020},
	pages = {1302--1313},
}

@article{Yao2023,
	title = {Broadband picometer-scale resolution on-chip spectrometer with reconfigurable photonics},
	volume = {12},
	journal = {Light: Science \& Applications},
	author = {Yao, Chunhui and Chen, Minjia and Yan, Ting and Ming, Liang and Cheng, Qixiang and Penty, Richard},
	year = {2023},
	pages = {156},
}

@article{Malone2023,
	title = {{DiffuserSpec}: spectroscopy with {Scotch} tape},
	volume = {48},
	journal = {Optics Letters},
	author = {Malone, Joseph D. and Aggarwal, Neerja and Waller, Laura and Bowden, Audrey K.},
	year = {2023},
	pages = {323--326},
}

@article{Lin2023,
	title = {High-performance, intelligent, on-chip speckle spectrometer using {2D} silicon photonic disordered microring lattice},
	volume = {10},
	journal = {Optica},
	author = {Lin, Zhongjin and Yu, Shangxuan and Chen, Yuxuan and Cai, Wangning and Lin, Becky and Song, Jingxiang and Mitchell, Matthew and Hammood, Mustafa and Jhoja, Jaspreet and Jaeger, Nicolas A. F. and Shi, Wei and Chrostowski, Lukas},
	year = {2023},
	pages = {497--504},
}

@article{Coluccelli2016,
	title = {The optical frequency comb fibre spectrometer},
	volume = {7},
	journal = {Nature Communications},
	author = {Coluccelli, Nicola and Cassinerio, Marco and Redding, Brandon and Cao, Hui and Laporta, Paolo and Galzerano, Gianluca},
	year = {2016},
	pages = {12995},
}

@article{Yang2017,
	title = {Compact {Spectrometer} {Based} on a {Frosted} {Glass}},
	volume = {29},
	journal = {IEEE Photonics Technology Letters},
	author = {Yang, T. and Huang, X. L. and Ho, H. P. and Xu, C. and Zhu, Y. Y. and Yi, M. D. and Zhou, X. H. and Li, X. A. and Huang, W.},
	year = {2017},
	pages = {217--220},
}

@article{Wang2014,
	title = {Computational spectrometer based on a broadband diffractive optic},
	volume = {22},
	journal = {Optics Express},
	author = {Wang, Peng and Menon, Rajesh},
	year = {2014},
	pages = {14575--14587},
}

@article{Kurekci2023,
	title = {Single-{Pixel} {Multimode} {Fiber} {Spectrometer} via {Wavefront} {Shaping}},
	volume = {10},
	journal = {ACS Photonics},
	author = {Kurekci, Sahin and Kahraman, S. Suleyman and Yuce, Emre},
	year = {2023},
	pages = {2488--2493},
}

@article{Redding2014,
	title = {High-resolution and broadband all-fiber spectrometers},
	volume = {1},
	journal = {Optica},
	author = {Redding, Brandon and Alam, Mansoor and Seifert, Martin and Cao, Hui},
	year = {2014},
	pages = {175--180},
}

@article{Yi2021,
	title = {Integrated {Multimode} {Waveguide} {With} {Photonic} {Lantern} for {Speckle} {Spectroscopy}},
	volume = {57},
	journal = {IEEE Journal of Quantum Electronics},
	author = {Yi, Dan and Zhang, Yaojing and Wu, Xinru and Tsang, Hon Ki},
	year = {2021},
	pages = {0600108},
}

@article{Redding2016,
	title = {Evanescently coupled multimode spiral spectrometer},
	volume = {3},
	journal = {Optica},
	author = {Redding, Brandon and Fatt Liew, Seng and Bromberg, Yaron and Sarma, Raktim and Cao, Hui},
	year = {2016},
	pages = {956--962},
}

@article{Zhang2022,
	title = {Cascaded nanobeam spectrometer with high resolution and scalability},
	volume = {9},
	journal = {Optica},
	author = {Zhang, Jiahui and Cheng, Ziwei and Dong, Jianji and Zhang, Xinliang},
	year = {2022},
	pages = {517--521},
}

@article{Redding2013,
	title = {Compact spectrometer based on a disordered photonic chip},
	volume = {7},
	journal = {Nature Photonics},
	author = {Redding, Brandon and Liew, Seng Fatt and Sarma, Raktim and Cao, Hui},
	year = {2013},
	pages = {746--751},
}

@article{Facchin2024,
	title = {Determining intrinsic sensitivity and the role of multiple scattering in speckle metrology},
	journal = {Nature Reviews Physics},
	author = {Facchin, Morgan and Khan, Saba N. and Dholakia, Kishan and Bruce, Graham D.},
	year = {2024},
	pages = {500--508},
    volume = {6},
}

@article{Redding_noise_2014,
	title = {Noise analysis of spectrometers based on speckle pattern reconstruction},
	volume = {53},
	journal = {Applied Optics},
	author = {Redding, Brandon and Popoff, Sebastien M. and Bromberg, Yaron and Choma, Michael A. and Cao, Hui},
	year = {2014},
	pages = {410--417},
}

@article{Korsch2006,
	title = {On two-dimensional {Bessel} functions},
	volume = {39},
	journal = {Journal of Physics A: Mathematical and General},
	author = {Korsch, H J and Klumpp, A and Witthaut, D},
	year = {2006},
	pages = {14947},
}

@article{Hang2010,
	title = {Photonic bandgap fiber bundle spectrometer},
	volume = {49},
	journal = {Applied Optics},
	author = {Hang, Qu and Ung, Bora and Syed, Imran and Guo, Ning and Skorobogatiy, Maksim},
	year = {2010},
	pages = {4791--4800},
}

@article{Redding_allfiber_2013,
    author = {Redding, Brandon and Popoff, Sebastien M. and Cao, Hui},
    title = {All-fiber spectrometer based on speckle pattern reconstruction},
    journal = {Optics Express},
    volume = {21},
    pages = {6584--6600},
    year = {2013},
}

@article{Rayleigh1879,
    author = {Lord Rayleigh, F. R. S.},
    title = {Investigations in Optics, witlt special reference to the Spectroscope.},
    journal = {Philosophical Magazine},
    volume = {8},
    pages = {261--274},
    year = {1879},
}

@article{Uttam2020,
    author = {Paudel, Uttam and Rose, Todd},
    title = {Ultra-high resolution and broadband chip-scale speckle enhanced Fourier-transform spectrometer},
    journal = {Optics Express},
    year = {2020},
    volume = {28},
    pages = {16469--16485},
}

@book{goodmanSpeckle,
    author = {Goodman, Joseph W.},
    title = {Speckle Phenomena in Optics: Theory and Applications},
    edition= {2nd},
    publisher = {SPIE},
    address = {Bellingham, WA},
    year = {2020},
}

@misc{steck2025,
    author = {Steck, Daniel A.},
    title = {Rubidium 85 D line Data},
    howpublished = {\url{http://steck.us/alkalidata}} ,
    year = {2025},
    note = {Revision 2.3.4, 8 August},
}

@misc{perrella_twisted_2025,
	title = {Sensing with Twisted Light: Precision Measurement of Fractional Azimuthal Index to Determine Refractive Index},
	publisher = {arXiv},
	author = {Perrella, Christopher and Punse, Anil Aman and Zalogina, Anastasiia and Szydzik, Crispin and Lim, Megan and Boes, Andreas and Mitchell Arnan and Dunning, R. Kylie and Dholakia, Kishan},
	year = {2025},
	note = {arXiv:2508.19521},
}

@article{Marpaung2019,
    author = {Marpaung, David and Yao, Jianping and Capmany, Jos\'{e}},
    title = {Integrated microwave photonics},
    journal = {Nature Photonics},
    year = {2019},
    volume = {13},
    pages = {80--90},
}

@article{RomeroCortes2020,
    author = {Romero Cort\'{e}s, Luis and Onori, Daniel and Guillet de Chatellus, Hugues and Burla, Maurizio and Aza\~{n}a, Jos\'{e}},
    title = {Towards on-chip photonic-assisted radio-frequency spectral measurement and monitoring},
    journal = {Optica},
    pages = {434--447},
    volume = {7},
    year = {2020},
}

@article{metzger_harnessing_2017,
	title = {Harnessing speckle for a sub-femtometre resolved broadband wavemeter and laser stabilization},
	volume = {8},
	journal = {Nature Communications},
	author = {Metzger, Nikolaus Klaus and Spesyvtsev, Roman and Bruce, Graham D. and Miller, Bill and Maker, Gareth T. and Malcolm, Graeme and Mazilu, Michael and Dholakia, Kishan},
	year = {2017},
	pages = {15610},
}

@article{Zou2016,
    author = {Zou, Xihua and Lu, Bing and Pan, Wei and Yan, Lianshan and Stöhr, Andreas and Yao, Jianping},
    title = {Photonics for microwave measurements},
    journal = {Laser \& Photonics Reviews},
    volume = {10},
    pages = {711-734},
    year = {2016},
}

@article{schreck_laser_2021,
	title = {Laser cooling for quantum gases},
	volume = {17},
	urldate = {2026-01-13},
	journal = {Nature Physics},
	author = {Schreck, Florian and Druten, Klaasjan van},
	year = {2021},
	pages = {1296--1304},
}

@article{LIGO_2015,
year = {2015},
publisher = {IOP Publishing},
volume = {32},
pages = {074001},
author = {{The LIGO Scientific Collaboration \textit{et al.}}},
title = {Advanced LIGO},
journal = {Classical and Quantum Gravity},
}

@article{alnis_subhertz_2008,
	title = {Subhertz linewidth diode lasers by stabilization to vibrationally and thermally compensated ultralow-expansion glass {F}abry-{P}\'erot cavities},
	volume = {77},
	journal = {Physical Review A},
	author = {Alnis, J. and Matveev, A. and Kolachevsky, N. and Udem, Th. and H\"ansch, T. W.},
	year = {2008},
	pages = {053809},
}

@article{Ludlow2015,
	title = {Optical atomic clocks},
	volume = {87},
	journal = {Reviews of Modern Physics},
	author = {Ludlow, Andrew D. and Boyd, Martin M. and Ye, Jun and Peik, E. and Schmidt, P. O.},
	year = {2015},
	pages = {637--701},
}

@article{kefelian_ultralow-frequency-noise_2009,
	title = {Ultralow-frequency-noise stabilization of a laser by locking to an optical fiber-delay line},
	volume = {34},
	journal = {Optics Letters},
	author = {K\'ef\'elian, Fabien and Jiang, Haifeng and Lemonde, Pierre and Santarelli, Giorgio},
	year = {2009},
	pages = {914--916},
}

@article{davila-rodriguez_compact_2017,
	title = {Compact, thermal-noise-limited reference cavity for ultra-low-noise microwave generation},
	volume = {42},
	journal = {Optics Letters},
	author = {Davila-Rodriguez, J. and Baynes, F. N. and Ludlow, A. D. and Fortier, T. M. and Leopardi, H. and Diddams, S. A. and Quinlan, F.},
	year = {2017},
	pages = {1277--1280},
}

@article{dong_subhertz_2015,
	title = {Subhertz linewidth laser by locking to a fiber delay line},
	volume = {54},
	journal = {Applied Optics},
	author = {Dong, Jing and Hu, Yongqi and Huang, Junchao and Ye, Meifeng and Qu, Qiuzhi and Li, Tang and Liu, Liang},
	year = {2015},
	pages = {1152--1156},
}

@article{katori_optical_2011,
	title = {Optical lattice clocks and quantum metrology},
	volume = {5},
	journal = {Nature Photonics},
	author = {Katori, Hidetoshi},
	year = {2011},
	pages = {203--210},
}

@article{safronova_search_2018,
	title = {Search for new physics with atoms and molecules},
	volume = {90},
	journal = {Reviews of Modern Physics},
	author = {Safronova, M. S. and Budker, D. and DeMille, D. and Kimball, Derek F. Jackson and Derevianko, A. and Clark, Charles W.},
	year = {2018},
	pages = {025008},
}

@article{Grotti2018,
	title = {Geodesy and metrology with a transportable optical clock},
	volume = {14},
	journal = {Nature Physics},
	author = {Grotti, Jacopo and Koller, Silvio and Vogt, Stefan and Häfner, Sebastian and Sterr, Uwe and Lisdat, Christian and Denker, Heiner and Voigt, Christian and Timmen, Ludger and Rolland, Antoine and Baynes, Fred N. and Margolis, Helen S. and Zampaolo, Michel and Thoumany, Pierre and Pizzocaro, Marco and Rauf, Benjamin and Bregolin, Filippo and Tampellini, Anna and Barbieri, Piero and Zucco, Massimo and Costanzo, Giovanni A. and Clivati, Cecilia and Levi, Filippo and Calonico, Davide},
	year = {2018},
	pages = {437--441},
}

@article{Hilton2018a,
	title = {High-efficiency cold-atom transport into a waveguide trap},
	volume = {10},
	journal = {Physical Review Applied},
	author = {Hilton, A.P. and Perrella, C. and Benabid, F. and Sparkes, B.M. and Luiten, A.N. and Light, P.S.},
	year = {2018},
	pages = {044034},
}

@article{diddams_optical_2020,
	title = {Optical frequency combs: {Coherently} uniting the electromagnetic spectrum},
	volume = {369},
	journal = {Science},
	author = {Diddams, Scott A. and Vahala, Kerry and Udem, Thomas},
	year = {2020},
	pages = {eaay3676},
}

@article{chiu_continuous_2025,
	title = {Continuous operation of a coherent 3,000-qubit system},
	volume = {646},
	journal = {Nature},
	author = {Chiu, Neng-Chun and Trapp, Elias C. and Guo, Jinen and Abobeih, Mohamed H. and Stewart, Luke M. and Hollerith, Simon and Stroganov, Pavel L. and Kalinowski, Marcin and Geim, Alexandra A. and Evered, Simon J. and Li, Sophie H. and Lyu, Xingjian and Peters, Lisa M. and Bluvstein, Dolev and Wang, Tout T. and Greiner, Markus and Vuletić, Vladan and Lukin, Mikhail D.},
	year = {2025},
	pages = {1075--1080},
}

@article{Bianco2018,
    title = {Strategies for reducing speckle noise in digital holography},
    author = {Bianco, Vittorio and Memmolo, Pasquale and Leo, Marco and Montresor, Silvio and Distante, Cosimo and Paturzo, Melania and Picart, Pascal and Javidi, Bahram and Ferraro, Pietro},
    journal = {Light: Science \& Applications},
    volume = {7},
    year = {2018},
    pages = {48}
}

@article{Bates1982,
  title={Astronomical speckle imaging},
  author={Bates, Richard H. T.},
  journal={Physics Reports},
  volume={90},
  pages={203--297},
  year={1982},
}

@article{Fei2024,
    author = {Xia, Fei and Kim, Kyungduk and Eliezer, Yaniv and Han, SeungYun and Shaughnessy, Liam and Gigan, Sylvain and Cao, Hui},
    title = {Nonlinear optical encoding enabled by recurrent linear scattering},
    journal = {Nature Photonics},
    year = {2024},
    volume={18},
    pages={1067--1075},
}

@article{Kelley2024,
    author = {Kelley, Matthew J. and Shaw, Thomas J. and Valley, George C.},
    title = {Radio-frequency spectrometry based on laser speckle imaging},
    journal = {Journal of Optical Microsystems},
    year = {2024},
    volume={4},
    pages={024502},
}

@article{Knappe2004,
    author = {Knappe, Svenja and Shah, Vishal and Schwindt, Peter D. D. and Hollberg, Leo and Kitching, John and Liew, Li-Anne and Moreland, John},
    title = {A microfabricated atomic clock},
    journal = {Applied Physics Letters},
    year = {2004},
    volume={85},
    pages={1460--1462},
}

@article{Ebadi2021,
    author = {Ebadi, Sepehr and Wang, Tout T. and Levine, Harry and Keesling, Alexander and Semeghini, Giulia and Omran, Ahmed and Bluvstein, Dolev and Samajdar, Rhine and Pichler, Hannes and Ho, Wen Wei and Choi, Soonwon and Sachdev, Subir and Greiner, Markus and Vuletić, Vladan and Lukin, Mikhail D.},
	title = {Quantum phases of matter on a 256-atom programmable quantum simulator},
    journal = {Nature},
    year = {2021},
	volume = {595},
	pages = {227--232}
}

@article{Peters1999,
    author = {Peters, Achim and Chung, Keng Yeow and Chu, Steven},
	title = {Measurement of gravitational acceleration by dropping atoms},
	journal = {Nature},
    year = {1999},
	volume = {400},
    pages = {849--852}
}

@article{chen_high_2023,
author = {Chen, Hongzhou and Duan, Zhenyu and Guan, Chunying and Gao, Shan and Ye, Peng and Liu, Yan and Yang, Jing and Liu, Hongchao and Shi, Jinhui and Yang, Jun and Yuan, Libo},
title = {A high resolution compact all-fiber spectrometer based on periodic refractive index modulation},
journal = {Applied Physics Letters},
year = {2023},
volume = {122},
pages = {231101},
}

@article{deng_compact_2024,
author = {Deng, Guoliang and Xu, Yunlong and Cai, Rui and Zhao, Hong and Wu, Jie and Zhou, Hao and Zhang, Hong and Zhou, Shouhuan},
title = {Compact speckle spectrometer based on {CNN}-{LSTM} denoising},
journal = {Optics Letters},
year = {2024},
volume = {49},
pages = {6521--6524},
}

@article{french_speckle-based_2017,
author = {French, Rebecca and Gigan, Sylvain and Muskens, Otto L.},
title = {Speckle-based hyperspectral imaging combining multiple scattering and compressive sensing in nanowire mats},
journal = {Optics Letters},
year = {2017},
volume = {42},
pages = {1820--1823},
}

@article{guo_high-resolution_2024,
author = {Guo, Zhipeng and Zhang, Long and Dai, Daoxin},
title = {High-Resolution Spatial Speckle Reconstructive Spectrometer Based on Random Reflection},
journal = {2024 22nd International Conference on Optical Communications and Networks ({ICOCN})},
year = {2024},
volume = {},
pages = {1--3},
}

@article{kohlgraf-owens_transmission_2010,
author = {Kohlgraf-Owens, Thomas W. and Dogariu, Aristide},
title = {Transmission matrices of random media: means for spectral polarimetric measurements},
journal = {Optics Letters},
year = {2010},
volume = {35},
pages = {2236--2238},
}

@misc{kumar_performance_2025,
author = {Kumar, Bhupesh and Bruce, Graham D. and Negro, Luca Dal and Schulz, Sebastian A.},
title = {Performance limit of on-chip speckle spectrometers},
publisher = {arXiv},
year = {2025},
note = {arXiv:2510.09077},
}

@article{lee_reconstructive_2025,
author = {Lee, Dong-gu and Song, Gookho and Lee, Chunghyung and Lee, Chanseok and Jang, Mooseok},
title = {Reconstructive spectrometer using double-layer disordered metasurfaces},
journal = {Science Advances},
year = {2025},
volume = {11},
pages = {eadv2376},
}

@article{liang_polarization_2024,
author = {Liang, Junrui and Ye, Jun and Ke, Yanzhao and Zhang, Yang and Ma, Xiaoya and He, Junhong and Li, Jun and Xu, Jiangming and Leng, Jinyong and Zhou, Pu},
title = {Polarization transmission matrix enabled high-accuracy, large-bandwidth speckle-based reconstructive spectrometer},
journal = {Applied Physics Letters},
year = {2024},
volume = {124},
pages = {071111},
}

@article{liang_high-resolution_2025,
author = {Liang, Junrui and Li, Jun and Huang, Zhongming and He, Junhong and Guo, Yidong and Ma, Xiaoya and Ke, Yanzhao and Ye, Jun and Xu, Jiangming and Leng, Jinyong and Zhou, Pu},
title = {High-resolution miniaturized speckle spectrometry using fuse-induced fiber microvoids},
journal = {Photonics Research},
year = {2025},
volume = {13},
pages = {2654--2667},
}

@article{liew_broadband_2016,
author = {Liew, Seng Fatt and Redding, Brandon and Choma, Michael A. and Tagare, Hemant D. and Cao, Hui},
title = {Broadband multimode fiber spectrometer},
journal = {Optics Letters},
year = {2016},
volume = {41},
pages = {2029--2032},
}

@article{meng_multimode_2019,
author = {Meng, Ziyi and Li, Jianqiang and Yin, Chunjing and Zhang, Tian and Yu, Zhenming and Tang, Ming and Tong, Weijun and Xu, Kun},
title = {Multimode fiber spectrometer with scalable bandwidth using space-division multiplexing},
journal = {{AIP} Advances},
year = {2019},
volume = {9},
pages = {015004},
}

@article{sun_compact_2022,
author = {Sun, Qi and Falak, Przemyslaw and Vettenburg, Tom and Lee, Timothy and Phillips, David B. and Brambilla, Gilberto and Beresna, Martynas},
title = {Compact nano-void spectrometer based on a stable engineered scattering system},
journal = {Photonics Research},
year = {2022},
volume = {10},
pages = {2328--2336},
}

@article{tan_optical_2023,
author = {Tan, Henry and Li, Bingxi and Crozier, Kenneth B.},
title = {Optical fiber speckle spectrometer based on reversed-lens smartphone microscope},
journal = {Scientific Reports},
year = {2023},
volume = {13},
pages = {12958},
}

@entry{tian_single-shot_2025,
author = {Tian, Wenzhang and Chen, Hao and Zhang, Mingyuan and Chen, Zengqi and Tong, Yeyu},
title = {Single-Shot Integrated Speckle Spectrometer with Ultrahigh Bandwidth-to-Resolution},
publisher = {arXiv},
year = {2025},
note = {arXiv:2505.15166},
}

@article{wang_high_2023,
author = {Wang, Xin and Sun, Qi and Chu, Yushi and Brambilla, Gilberto and Wang, Pengfei and Beresna, Martynas},
title = {High resolution compact spectrometer system based on scattering and spectral reconstruction},
journal = {Optics Letters},
year = {2023},
volume = {48},
pages = {1466--1469},
}

@article{zhang_scalable_2025,
author = {Zhang, Zimeng and Xiao, Shumin and Song, Qinghai and Xu, Ke},
title = {Scalable on-chip diffractive speckle spectrometer with high spectral channel density},
journal = {Light: Science \& Applications},
year = {2025},
volume = {14},
pages = {130},
}

@article{zhang_camera-free_2025,
author = {Zhang, Mingyuan and Tian, Wenzhang and Lu, Kaihang and Chen, Hao and Zhou, Wu and Tong, Yeyu},
title = {A camera-free and picometer-scale resolution few-mode fiber spectrometer},
journal = {{APL} Photonics},
year = {2025},
volume = {10},
pages = {076109},
}

@article{cen_microtaper_2023,
author = {Cen, Qingqing and Pian, Sijie and Liu, Xinhang and Tang, Yuwei and He, Xinying and Ma, Yaoguang},
title = {Microtaper leaky-mode spectrometer with picometer resolution},
journal = {eLight},
year = {2023},
volume = {3},
pages = {9},
}

@article{hartmann_broadband_2020,
author = {Hartmann, Wladick and Varytis, Paris and Gehring, Helge and Walter, Nicolai and Beutel, Fabian and Busch, Kurt and Pernice, Wolfram},
title = {Broadband Spectrometer with Single-Photon Sensitivity Exploiting Tailored Disorder},
journal = {Nano Letters},
year = {2020},
volume = {20},
pages = {2625--2631},
}

@article{piels_compact_2017,
author = {Piels, Molly and Zibar, Darko},
title = {Compact silicon multimode waveguide spectrometer with enhanced bandwidth},
journal = {Scientific Reports},
year = {2017},
volume = {7},
pages = {43454},
}

@article{li_inverse-designed_2023,
author = {Li, Yuan and Zhang, Zunyue and Wang, Yi and Yu, Yue and Zhou, Xuetong and Tsang, Hon Ki and Sun, Xiankai},
title = {Inverse-Designed Linear Coherent Photonic Networks for High-Resolution Spectral Reconstruction},
journal = {{ACS} Photonics},
year = {2023},
volume = {10},
pages = {1012--1018},
}

@article{wan_high-resolution_2015,
author = {Wan, Noel H. and Meng, Fan and Schröder, Tim and Shiue, Ren-Jye and Chen, Edward H. and Englund, Dirk},
title = {High-resolution optical spectroscopy using multimode interference in a compact tapered fibre},
journal = {Nature Communications},
year = {2015},
volume = {6},
pages = {7762},
}

@article{yi_high-resolution_2022,
author = {Yi, Dan and Tsang, Hon Ki},
title = {High-Resolution and Broadband Two-Stage Speckle Spectrometer},
journal = {Journal of Lightwave Technology},
year = {2022},
volume = {40},
pages = {7969--7976},
}

@article{zhang_compact_2021,
author = {Zhang, Zunyue and Li, Yuan and Wang, Yi and Yu, Zejie and Sun, Xiankai and Tsang, Hon Ki},
title = {Compact High Resolution Speckle Spectrometer by Using Linear Coherent Integrated Network on Silicon Nitride Platform at 776 nm},
journal = {Laser \& Photonics Reviews},
year = {2021},
volume = {15},
pages = {2100039},
}

@article{zhang_miniature_2024,
author = {Zhang, Yangxi and Zhang, Sheng and Wu, Hao and Wang, Jinhui and Lin, Guang and Zhang, A. Ping},
title = {Miniature computational spectrometer with a plasmonic nanoparticles-in-cavity microfilter array},
journal = {Nature Communications},
year = {2024},
volume = {15},
pages = {3807},
}

@article{redding_using_2012,
author = {Redding, Brandon and Cao, Hui},
title = {Using a multimode fiber as a high-resolution, low-loss spectrometer},
journal = {Optics Letters},
year = {2012},
volume = {37},
pages = {3384--3386},
}

\end{document}